# Photolithography Control System : A Case Study For Cyber-Physical System

Youbao Zhang, Huijie Huang

Shanghai Institute of Optics and Fine Mechanics, Chinese Academy of Sciences, Shanghai 201800, China

# Abstract

Photolithography control system (PCS) is an extremely complex distributed control system, which is composed of dozens of networked microprocessors, hundreds of actuators, hundreds of thousands of sensors, and millions of lines of code. Cyber-physical system (CPS), which deeply merges computation with physical processes together, copes with complex system from a higher level of abstraction. PCS is a representative CPS. This work points out that thinking under the framework of CPS, which includes holistic perspective, model-based design, hardware/software co-design and continuous integration, could solve the issues presented in the current PCS. Although the traditional embedded system approach and the CPS approach would be coexisting in the PCS for a long time, the CPS approach is definitely the future of the PCS development.

# 1 Introduction

Photolithography is a process of transferring pattern on a mask to the surface of a silicon wafer with nanometer precision. Lithography machine is the critical equipment for the photolithography process, and it is the heart of the modern semiconductor fabrication plants. As the modern industry's most brilliant pearl, lithography machine is one of the most sophisticated optomechatronical machines human ever created, and only a few companies, such as ASML, Nikon, Canon, and SMEE, could manufacture it. It is remarkably complicated, and a modern lithography machine can be broken into more than 50,000 parts [1]. It is extremely precise, and the state-of-the-art line width is about 7 nm, exposing from the cutting-edge EUV machine [2]. It is incredibly expensive, and it costs hundreds of millions of dollars, and even billions of dollars for the cutting-edge models, expensive more than a modern Boeing airplane. It is estimated that the control system, including hardware, software and algorithm, almost cost one third of the whole machine, and it also occupies one third the volume of the entire machine. Photolithography control system (PCS) is composed of a cluster of complex embedded systems, which could be regarded as a cyber-physical system (CPS) from a higher level of abstraction.

CPS, which deeply merges computation with physical processes together, copes with complex system from a higher level of abstraction, trying to make system design from a scientific approach instead of the traditional ad hoc approach, grasping the system as a whole from a systematic way, and making it possible to develop more dependable systems [3]. The higher level of abstraction has to embrace physical dynamics and computation in a unified way.

This paper aims at presenting a complex distributed real-time control system for CPS research. Meanwhile, the paper is trying to employ the promising research result of the CPS to tackle the issues presented in the PCS. This paper firstly summarizes CPS's development and achievements, and outlines how to develop systems and projects under the framework of CPS in Section 2. Then the PCS is comprehensively reviewed from a perspective of traditional embedded system in Section 3. In Section 4, the paper provides the means of transfering the traditional embedded system into the framework of CPS. In Section 5, a discussion is presented to describe the mutual influence and contributions between PCS and the CPS. Finally, a conclusion is given that the CPS is the future of the PCS development. Some references are cited from the published academic journals and books, and there are considerable contents are empirical and practical, or learned from the senior industry peers.

# 2 CPS development and achievements

The relationship between cyber and physical process is evolving with the progress of technologies. In the first beginning, the system was entirely composed of mechanical parts, and there was no cyber at all. As the emergence of computers, the connection between computation and physical system happened, and the computers were used for system control, which was mainly open loop control in the initial stage. With the miniaturization of computer systems, the system further fused computation and physical process, and the concept of embedded system has emerged [4]. The early embedded systems were mostly standalone, and they were usually small, easy porting and configuring, without or with only limited real-time requirements. Later, there were networked complex embedded systems, which need a great effort to cope with based on the existing technologies and methodologies. Following the Moore's Law, as the further development of semiconductor technology and micro-electronics, computation was ubiquitous, the concept of CPS was emerged, and the computation and physical process are deeply merged than ever before [5]. CPS revolutionizes our interaction with the physical world, and scientists and engineers expect the new concept would solve the existing issues in the traditional embedded system.

After decades of development, the achievements in the CPS research are promising [6].
1. The CPS concept, which treats the traditional embedded system from a holistic perspective, was introduced. The difference between the CPS and the traditional embedded system was point out, the characteristics and challenges of CPS were summarized, and the research areas and directions were suggested.
2. The challenges for the development of CPS were identified. To begin with, Edward Lee points out that the time properties lacking in the core abstraction of computing impeded the fully development of CPS [7, 8]. Moreover, time predictable computer architecture is still needed to be researched [9]. Last but not least, tailored for CPS operating system and CPS-friendly general purpose programming language are on the high wish list.
3. Scientists and engineers are exploring formal methodology for CPS research [10], and they are trying to employ the model-based approach to replace the traditional ad hoc approach [11]. The demanding for systematic modelling language has inspired a new wave of research on system modelling languages, such as POOSL, SysML, and SystemC [12, 13].
4. CPS applications are extensively explored during the past decades. CPS are applied to an extremely wide-range of application areas, and they are spanning different scales. These application areas include but not limited to vehicular system and transportation, aerospace and air-traffic management, defense, renewable energy, national power-grid system, environmental monitoring, manufacturing and industrial process control, robotics, medical and health-care system, networking system, cloud computing and data centers [14, 15].
5. Miscellaneous aspects for CPS are broadly studied by the researchers, such as dependability, resiliency, security and privacy, artificial intelligence in CPS, humans in the loop, and verification and validation.

In summary, CPS research has achieved encouraging results; however, CPS challenges have not been conquered yet, systematic methodologies are still in development, influential research results have not produced yet, and CPS research is still in its infancy. It needs time to overcome the challenges, develop formal design methodologies, and deal with a variety of issues. However, CPS projects are confronted every day; therefore, we have the following suggestions.

1. Learn experience from the state-of-the-art embedded systems, and merge those experience into the CPS framework.
2. Make incremental improvement for the existing embedded systems, which is the most feasible solution before conquering the challenges of the CPS.
3. If possible, migrate the existing projects to the framework of CPS.
4. Taking advantages of multiple discipline cutting-edge technologies, design and deployment new projects under the framework of CPS if possible.

# 3 Photolithography control system (PCS)

PCS is an extremely complex distributed control system, which is composed of dozens of networked microprocessors, hundreds of actuators, hundreds of thousands of sensors, and millions of lines of code. The system is physically integrated into a tight compact volume. To simplify the development and integration, the control system is divided into dozens of subsystems, with a centralized workstation tightly controlling all the subsystems. To operate the system, various subsystems have to be controlled in a coordinated and synchronized way. To facilitate the following expression, the abbreviations of the subsystems and other terms were introduced in Table 1.

Table 1 Abbreviations of subsystems and other terms

| Abbreviation | Subsystems and other terms |
|---|---|
| RS | Reticle stage subsystem |
| WS | Wafer stage subsystem |
| RH | Reticle handling subsystem |
| WH | Wafer handling subsystem |
| IPC | Illumination and projection controlling subsystem |
| AL | Alignment subsystem |
| LS | Focusing and levelling sensing subsystem |
| SC | Synchronous controlling subsystem |
| SH | Safety handling subsystem |
| VS | Variable slit component |
| HSSL | High speed serial link |

## 3.1 Physical process

The fabrication of an integrated circuit requires a variety of physical and chemical processes, such as prepare wafer, coat with photoresist, prebake, align and expose, develop, etch and implant, strip resist, and packaging, which are illustrated in detail in Figure 1 [16, 17]. Exposure is the most critical step, which determines the resolution of the lithography to a large extent, and lithography machine is the critical equipment for exposing.

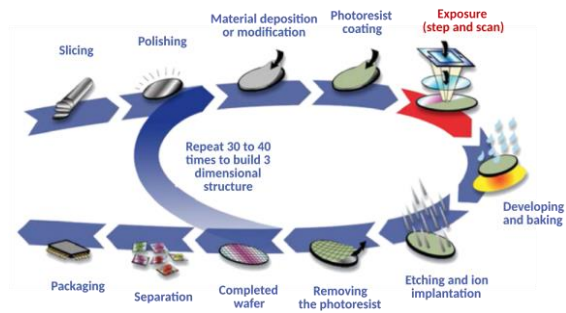

Figure 1 Photolithography workflow (courtesy of Umut Uyumaz)

The primary purpose of lithography machines is to project the nanoscale patterns in the reticle onto the wafer, and the state-of-the-art pattern dimension is 7nm. There are two kinds of tools: steppers and scanners. A wafer stepper fully exposes the complete reticle, while a scanner exposes the complete reticle through scanning, i.e. moving the reticle and wafer in opposite directions in a synchronized way, as shown in Figure 2 [18]. The scanning technology reduces the complexity and cost of optics and allows for larger integrate circuit exposure areas, but it dramatically increases the complexity of the control system [19].

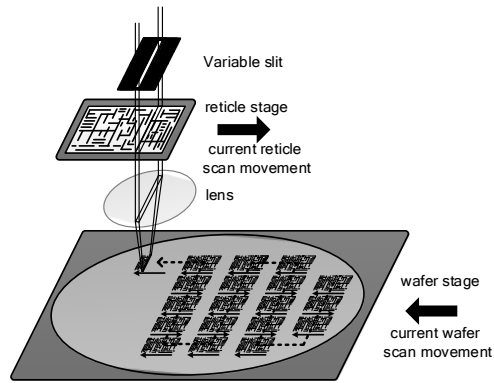

Figure 2 Basic principle of scanners (courtesy of Raymond Frijns)

The exposure workflow is shown in Figure 3. The prepared wafer is firstly loaded onto the wafer stage. Then the alignment system is activated to align the wafer to the reticle, which is the critical step since a wafer can be exposed with up to 30 image layers with extremely precise and repeated overlay. After the alignment, the wafer is further focused and levelled to guarantee the photoresist surface height and tilting fitting within the narrow depth of focus. Exposing is the core physical process, and scanning is the most important action, which requires accurately synchronously moving of the RS, WS, and VS, while dose control generates the correct exposure dose. Finally, the wafer is unloaded for post-processing.

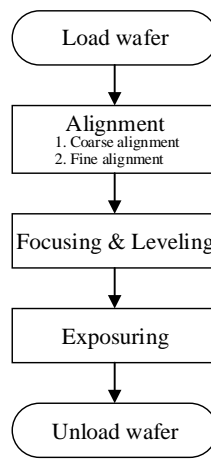

Figure 3 Photolithography exposing workflow

To increase the throughput, measuring and exposing are intentionally separated through employing the dual-stage technology, wherein one stage is used to measure to-be-exposing wafer's surface, while the other stage is simultaneously used for exposing the working wafer [20]. When both exposing and measuring is finished, the two stages switch positions. There are considerable concurrency physical processes like the dual-stage technology in lithography tools.

## 3.2 Architecture

Partitioning and hierarchy are two powerful strategies for handling the complexity of PCS. On one hand, the control system is partitioned into dozens of subsystems, and the main subsystems are including RH, WH, RS, WS, IPC, SC, AL, LS, and SH., which are connected to the central workstation through both Ethernet and RS232/RS485. Ethernet is mainly used for data transferring, and RS232/RS485 is primarily used for trouble shooting and firmware downloading. Sync Bus are used to synchronize the action of the participated subsystems for the critical scanning and exposing actions. The control system architecture is illustrated in Figure 4.

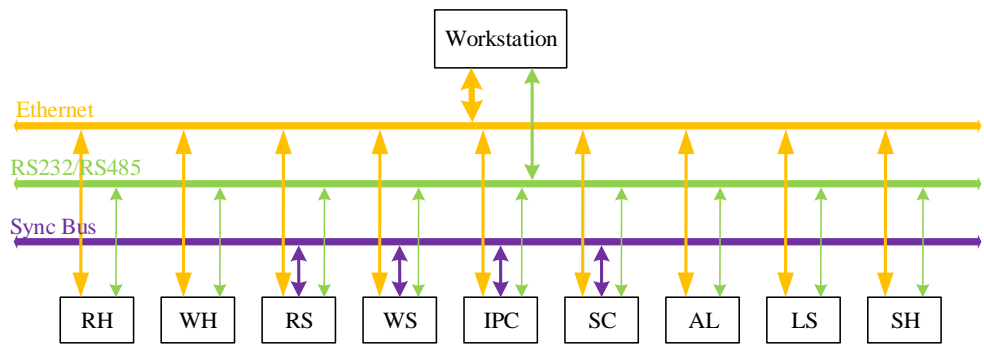

Figure 4. PCS architecture

On the other hand, the PCS is layered to simplify the development and manipulation, as shown in Figure 5. The workstation is on the top of the hierarchy for the high-level control, and the subsystems and below are all low-level control. For an individual subsystem, there is at least one CPU, and there are also co-processors such as DSP, FPGA, ARM, MCU and SoC. The I/O interface layer directly interact with the sensors and actuators in the lowest layer. The layered framework drastically simplifies the system design and programming development.

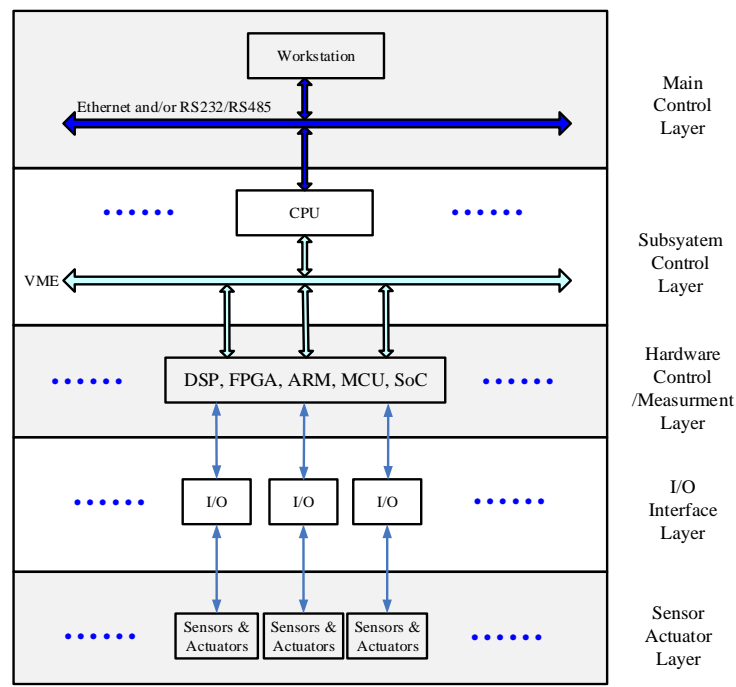

Figure 5 Hierarchy control architecture

## 3.3 Heterogeneous

As the most complicated machinery ever created, processors used in the PCS have a great diversity of architectures. General purpose processors are applied for general scenarios, while high-performance special purpose processors are designed for specific applications with high predictability. In fact, there are a variety of processors are adopted according to specific application, such as SPARC, X86, PowerPC, ARM,

FPGA, and DSP. For instance, the dose control board (DCB) is designed combined FPGA and DSP processors, and the FPGA processor is used for VME bus protocol and real-time signal acquiring, and the DSP processor is used for algorithm and data processing. The heterogeneity of processors brings the great diversity of operating systems and programming languages. Table 2 shows the processor architectures, the corresponding operating system and language support, and the main application scenarios.

Table 2 Processors architecture used in the PCS

| Processor architecture | OS support | Language | Application |
|---|---|---|---|
| SPARC | Solaris/Linux | C/C++, Python | High-level machine control |
| X86 | Windows | C/C++, matlab, Python | Metrology, Monitoring |
| PowerPC | VxWorks | C/C++ | Low-level real-time control |
| ARM | Linux/ucos etc. | C/C++ | General low-level control |
| FPGA | N/A | VHDL/Verilog | Bus Protocol, Logic control, Motion Control |
| DSP | N/A | C/C++ | Algorithm, Motion Control |

In addition to the processors listed in the above table, programmable logic controller (PLC) and application specific integrated circuit (ASIC) are employed in the control system. The safety system is monitored with a dedicated PLC, which isolated from the main control system, making the security more dependable. The ASIC chips, such as CCD, CMOS, and high-performance AD/DA. are also applied in the control system.

Since different processor types and different programming languages use different representations for the data they manipulated, it is a main challenge to interconnect these heterogeneous system, and many technologies are involved. For the hardware aspect, $I^2C$ and/or SPI bus are used in the chip level; backplane, UART or HSSL. are applied in the printed circuit board level; Ethernet, CAN bus, or RS485 are employed in the system level. For the software aspect, databases and customized middleware, such as communication networking component and open database connectivity component, are applied for software integration.

## 3.4 Real-time system

## Requirements & Design Principle

Since time is money for the modern semiconductor fabrication plants, engineers do their best to increase the overall throughput through enhancing the speed of a series of actions during lithography process. Actions are classified based on their time requirements. Some actions need to be rapidly responded, such as illumination mode setting, variable attenuation plate adjusting, numerical aperture setting, uniformity compensation plate adjusting, wafer loading and unloading, and if these actions do not complete in the scheduled time, the throughput will decrease; Some actions need to be tightly synchronously responded, otherwise, the wafer would be failed to expose. The most representative action is dose control, which involves the synchronous scanning of wafer stage, reticle stage and the variable slit, the real-time adjust of variable attenuation plate, the synchronous firing of the laser, the individual pulse energy synchronously acquiring, and the leveling and focusing, wherein the actions are concurrency, and the synchronous time error is significantly tight. The PCS finishes the above-mentioned actions through a variety of means: i) optimizing hardware resource deployment and task scheduling through real-time operating system for rapidly response to actions; ii) enabling interrupting mechanism for real-time response; iii) using hardware-based synchronous bus to complete the coordination of multiple components for hard real-time tasks; iv) employing high speed communication means whenever it is possible.

Based on field practice, the following design principles for lithography real-time control system are summarized:
1. Fully understand and differentiate the hard, firm, soft real-time, and non-real-time requirements, and dealing with them accordingly [21];
2. Make control loop as close as possible to the sensors and actuators, reducing the control loop hierarchy as much as possible；
3. Resort to hardware solutions if it is hard to implement the real-time requirements with software approach;
4. For the critical tasks with hard real-time requirements, directly implement with hardware, employing the cutting-edge technologies, such as DSP, FPGA, ASIC, multiprocessor and multi-core technologies [22].

As a rule of thumb, the boundary between software and hardware implementation for real-time requirements is on millisecond level. If the response time requirement is less than one millisecond, hardware implementation is the first choice; if the response time requirement is large than 50 milliseconds, software implementation is possible; and if the requirement is between 1 and 50 milliseconds, the implementation should be determined based on the application scenarios. Table 3 shows the time frame in photolithography machine tools.

Table 3 Time frame in photolithography machine tools

| Action/Operation | Time Frame |
| --- | --- |
| Time delay (from trigger to acquisition) for Pulse energy acquisition | Approximately 10μs |
| High-performance servo period | Dozens of microseconds, typically 20μs |
| Excimer laser pulsing period | Hundreds of microseconds |
| General servo period | Hundreds of microseconds, typically 400μs |
| Accelerating time for variable slit | Dozens of milliseconds |
| Mean die exposure time | Dozens of milliseconds |
| Illumination mode setting time | Several seconds, typically 3s |

| General mechanical setting time | Several seconds |
| Mean wafer exposure time | Dozens of seconds |
| Mean LOT exposure time | Several minutes |

## Global time base

In a distributed control system, a global time base should be introduced to guarantee a consistent behavior, such as causal order, temporal order, and delivery order etc., hence, all the events have a global time stamp. As a complex distributed control system, PCS involves considerable processes and actions, which need to be based on a unique global time base to guarantee the normal operations of the complex processes. The unified timestamp is used for reporting data, events, and alarms. In semiconductor fabrication plants, time synchronization in lithography tools is implemented through NTPv3 protocol, which guarantees a unified global time base in a lithography tool, among different tools in a fab, and even among different semiconductor fabrication plants across a wide area [23].

## Central Master Synchronization

The real-time performance requirements for lithography processes are significantly strict, and the real-time control by the software alone is unsatisfactory, especially for the synchronous scanning and the dose control in scanners. Therefore, a synchronous control system and a synchronous mechanism are specially designed. The synchronous control system is composed of a master synchronous control board (MSB), a synchronous bus, and a number of slave synchronous control boards (SSB), shown in Figure 6. The synchronous bus connects the master board and the slave boards, and when there is a synchronous requirement, the master board triggers signal to the synchronous bus, and the slave board which received the synchronous signal responds to the synchronous requirement, and realize the synchronous action. The real-time synchronous control based on hardware level is achieved with the synchronous bus. To decrease delay caused by signal transferring, long-distance synchronization could be implemented with optical fiber communication channels.

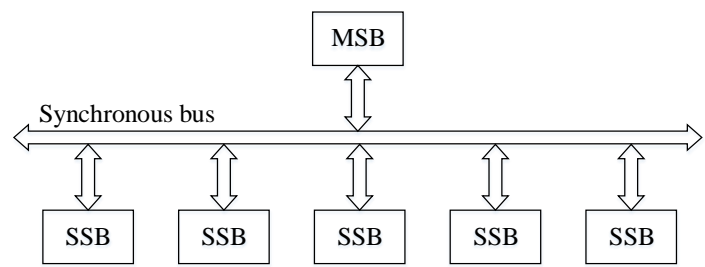

Figure 6 The synchronous control system

## 3.5 Software

## Overview

Modern equipment and devices are increasingly dependent on software, and lithography tool is a complex software-intensive system, controls and measurements being mainly driven by software. The lithography control software is designed such that mechanics and optics are fully utilized, and it is as complex as a modern computer operating system. There are more than 35 million lines of code in the control software in comparison with about 17 million lines of code for Linux kernel (version 4.20, estimated using David A. Wheeler's 'SLOCCount') [24] and about 45 million lines of code for Microsoft Windows XP [17]. As the increasing functional requirements, the processes are more complicated, the algorithm complexity is raising, the lines of code are dramatically increasing, and the software is getting incredibly complex.

Different from general purpose software, the lithography software has unique characteristics, which include: i) tightly coupled with hardware system; ii) deeply fused with photolithography physical processes; iii) distributed on a variety of CPU architectures and operating systems; iv) considerable actions are long time delayed processes, which mainly influence graphics user interface (GUI) design. To simplify the complex lithography control software, divide and conquer strategy is employed, and the concerns are separated into a series of simple scenarios. Layering and modularization, which are corresponding to the vertical and horizontal partitions, respectively, are common methods. Besides, configurable design simplified the software design from a holistic perspective.

## Software layering

In lithography tools, software is separated into several layers according to the hardware framework as shown in Figure 5, and each layer has a specific role and responsibility. The bottom layer is the sensors and actuators, which are the fundamental elements of the control system. Then, the I/O interface layer is dealing with the interface between the bottom layer and the embedded processors. The next layer is the hardware control and measurement layer, which implemented the fundamental control loop, and the most real-time processing is located here. The following layer is the subsystem control layer, which is responsible for the functions of the subsystems. The top layer is the main control layer, which implements high-level control, coordinating the entire system as a whole.

## Modularization

Layering partially simplifies the lithography control software, and modularization further facilitates the software development by decomposing the large software into smaller independent segments with well-defined interfaces. There are general functional modules and specialized modules in the lithography control software. For the general functional modules, there are communication networking module, trace module, database module, machine constant module, error handling and exception processing module, and diagnostic module; for the specialized modules, there are dose control module, levelling and focusing module, alignment module, measuring and calibration module, scanning and exposing module, and safety monitoring module. Modularization makes collaboration in parallel feasible for the complex lithography control software.

## Configuration

Configuration is required for complex software. Without configuration, any changes to the software would need a new compiling, which is considerable time consuming. For instance, recompiling the entire lithography control software would take more than thirty minutes, even hours. The flexibility and applicability are quite limited without configuration. Registry on Windows operation system and configuration files on Linux operating system are representative configuration means for operating systems.

Lithography control software is a complex industrial application software, which complexity is almost equivalent to an operating system, and it is a highly configurable software system. The configuration items include but not limited to feature toggle, simulation-level setting, parameters setting for hardware, and huge volumes of machine constants. The configuration is implemented through a series of configuration files and a dedicated database, which fulfill the requirements of flexibility and fast reconfiguration.

## 3.6 System Integration

## Overview

System integration aims at functional and physical process implementation, ensuring all the component parts work as a whole, and it is the most critical step for the project progress [25]. For the complex lithography system, divide and conquer strategy is employed. The system is firstly simplified through partition, isolation, and segmentation, and then after the individual component is finished, they are integrated through composition and synthesis, making the system act as a coordinated whole. The process of the project implementation is like a V-model in software development as shown in Figure 7. As a rule of thumb, system integration almost takes about more than one third of the project, and it directly determines the progress of the project. The control system integration, including hardware, firmware and software integration, are along with the whole process of the system integration.

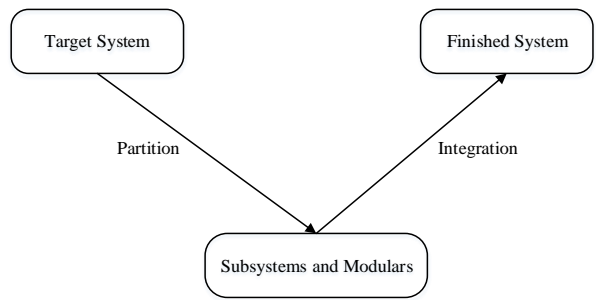

Figure 7 V-model of system integration

Lithography system integration is hierarchically implemented. In the system level, the lithography system is integrated by dozens of subsystems as shown in Figure 8. For an individual subsystem level, the integration includes optomechanical integration, measuring system integration, hardware and software integration, and physical process integration etc. Hardware and software integration will be described in detail in the following sections, respectively.

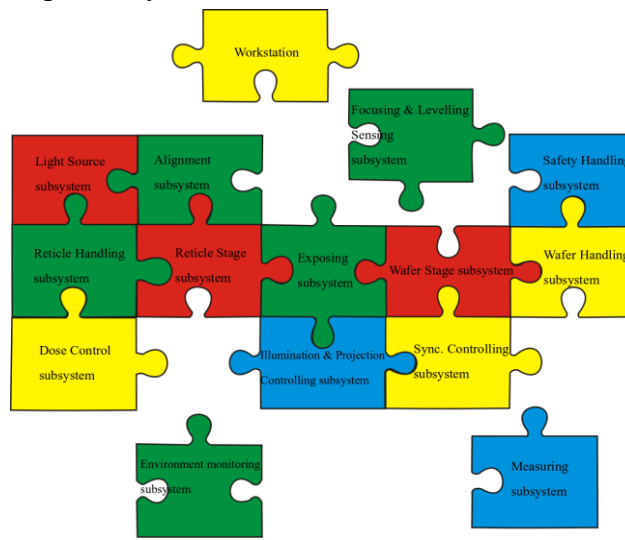

Figure 8 System integration based on subsystems

## Hardware integration

From transistors to chips, from discrete elements to printed circuit boards, to racks, to cabinets, and eventually to hardware systems, hardware integration, like constructing a pyramid, integrates the basic elements into a complicated system. Photolithography hardware integration hierarchy is shown in Figure 9. High-density integration is the trend for hardware system integration, and the integration level is increasing as the technology progress. For instance, the previous system-on-board function possibly implement on a key integrated circuit, which is so called system-on-silicon, and a function implemented with a cabinet may be implemented with a printed circuit board with a key integrated circuit chip. The integration capacity information is summarized in Table 4. For increasing hardware system integration in lithography tools, ASICs，FPGAs，and customized PCBs and racks. are extensively employed.

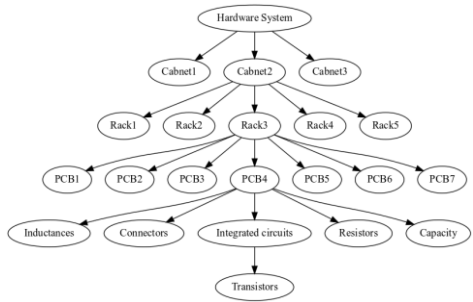

Figure 9 Hardware integration hierarchy

Table 4 Integration capacity

| Integration Form | Integration Level | Integration Layer |
|---|---|---|
| Integrated Circuit (IC) | Millions or billions of transistors | Element |
| Printed Circuit Board (PCB) | Hundreds of elements, including ICs, connectors, discrete elements | Component |
| Rack | Dozens of PCBs | Subsystem |
| Cabinet | Several Racks | System |

## Software integration

In lithography system, software integration, which involves interface, wrapping, middleware, binding, component granularity, glue codes, testing, is the main thread of the system integration, and it is the first phase before any stage of the system integration. Generally speaking, software integration is the most systematic effort, and it links all the parts of the system. Process integration, which means a physical process implementation, has been merged into the software integration, and it is actually software integration in a higher functional level. Compared to the general software, the scale and the complexity of the lithography software are astonishing, which are almost equivalent to an operating system. The software is integrated hierarchically, and from bottom to top is as following: unit-level integration, modular integration, subsystem integration, and system integration. Such complex software needs to be extensively tested at every integration level, and verification and validation is the most important procedure during software integration, guaranteeing the correctness and reliability.

## 3.7 Dependability

Dependability is the most fundamental requirement for lithography tools. Time is money, and down-time is very costly since the throughput is decreasing. The dependability is directly related to the profit of a fab, and the poor dependability would be compensated by significant expenditure. For the PCS, on the top of design and analysis, dependability is guaranteed through a variety of means.

In the lithography machine, there are sophisticated mechanism and methods to handling exceptions, errors, and recoveries. Lithography tools involves thousands of sensors and actuators, and tens of thousands of parts, it is not uncommon for failures and errors.

Some failures and errors are able to automatically recovery through predefined procures, which are mainly completed with software; Some failures and errors are handled by field engineers. It is the state-of-the-art that any industrial strength embedded system must have the capability to detect or mask the failure of its sensors and actuators, which must be tested by fault-injection experiments, either software-based or physical.

As a software-intensive system, the software dependability is the most important，and currently the main method to guarantee the software dependability is intensively testing. Although certification is a formal approach to ensure dependability，it is not applicable for lithography tools, since unlike airplanes, a model can last dozens of years [26], lithography tools model, however, following the Moore's Law, updates rapidly, and every model is about less than hundred, even less than ten, and in the extreme case, only one prototype is manufactured. The certification process and the reward is significant imbalance. Therefore, testing is indispensable, and no software system can be regarded as dependable if it has not been extensively tested. The software quality is assured by a variety of testing, such as unit testing, white-box testing, black-box testing, integration testing, and system testing.

# 4 Transfer PCS to CPS

Since there are a variety of issues in the traditional PCS, new abstraction system or methodology must be employed when facing higher performance of requirements and more complex situations. CPS would be the most suitable solution.

## 4.1 Issues in current PCS

First of all, the existing control system lacks a holistic perspective, and hardware design, software design and process control are virtually isolated. Software design is separated from hardware design, and it is often serial to hardware design. Instead of balancing across the whole system, real-time tasks are loaded on either a specific software layer or a specific hardware layer, or just burdened on the programmers. Well-defined design workflow is rarely formed.

Secondly, a systematic model is absent for the full control system analyzing, simulation, verification and validation. For the optical system, there is a specialized research direction called computational lithography, which tries to improve the resolution through modelling the entire optical system. For the mechanical system, the full system model could be built for further analyzing and simulation, such as static analysis, transient analysis, modal analysis, and fatigue analysis. But for the control system, only partial models could be built in a limited domain or in a local small scale, and the full control system modelling and simulation yet remains a daunting dream.

Thirdly, a systematic theory is yet missing for system integration, which mostly relies on ad hoc methods. The individual components are simply put together, making the system work somehow. There are considerable surprises when those components come together. The functions and the performance are then tested, and the correctness of system integration is ensured by intensively testing. Most time the integration is a long and uncertain process, and it is hardly to predict how long it will take. Transforming system integration from a high-risk practice into a science-based engineering discipline is a significant challenge.

Last but not least, the control system barely meets the increasing requirements for higher precision, higher throughput, and more complex control logic in the next generation lithography tools development. For instance, Flexray, a programmable illumination technology from ASML, which could freely adjust the pupil shape, involves thousands of micro-mirrors to be controlled, which is a formidable task with the present technology.

## 4.2 Solutions to the existing issues

CPS is the most suitable solution to the issues of the traditional embedded system. There are systematic methodologies for the CPS design, such as holistic perspective, model-based design, hardware/software co-design, and continuous integration, which could address the issues presented in the current PCS.

## Holistic perspective

Holistic perspective is a systematic high-level thinking, giving full control of the whole system [27, 28]. Partition is a strategy for simplifying problems, while holism is the natural property of the natural world. Since a system shows a total behavior, holistic thinking allows the system engineers to focus on the integration of all the components as a coherent and effective system. There are many advantages of the holistic perspective: i) achieve the optimal allocation of resources in the entire system, and implement optimal layering and modular partition; ii) implement optimal close loop control in the system which involves multiple loops; iii) holistic approach is the basis for model-based design and hardware/software co-design, and only thinking from a holistic perspective, the best results be could achieved. Holistic thinking is the fundamental feature of the CPS concept.

Treating the PCS as a CPS, and thinking under the framework of CPS from a holistic perspective, would benefit the system design and system integration. For instance, dose control is the most important function of the lithography tools, and it coordinates multiple components working together. Therefore, in order to achieve the best performance, the dose control system should be designed based on a holistic approach. Taking another example, safety must be taken into account from a holistic perspective, and the interlock system design must be designed from a holistic approach in the lithography safety system.

# Model-based design

Development of complex systems requires the creation of models, and it is the formal methodology. Models are not only used to predict and control the behavior of complex systems, but also to help us understand the systems better. Ad hoc approach could accumulate abundant of experience and lessons, but those are scattered knowledge, which are empirical and limited; whereas model-based approach could accumulate systematic knowledge in a specific domain, which could be further developed and extended. The benefits of the model-based design include architectural exploration, iterative optimization, performance analysis, algorithm determination and proof, early software development, early integration, and detection of design defects. The functions of model are illustrated in Figure 10 [29, 30, 31].

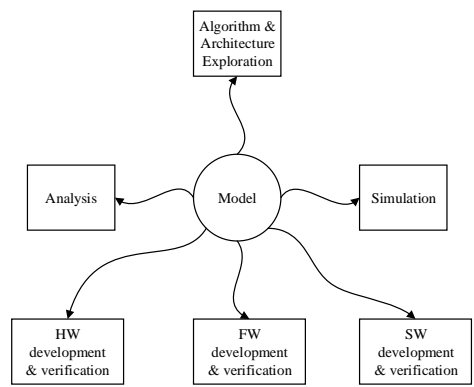

Figure 10 The functions of model

As the increasing complexity of models, systematic modelling languages have emerged, such as POOSL and SystemC. Both POOSL and SystemC are system-level modelling language, including both software and hardware. SystemC, which provides a notion of timing as well as an event driven simulations environment, allows the description and integration of complex hardware and software components in the C++ language, implementing hardware/software co-design and co-simulation in one environment [32]. In PCS, motion control system is the representative application for systematic modeling with system-level modelling language. Low-level hardware model could be created on the register-transfer level with SystemC language; and application level model could be created on the transaction-level with the same modelling language. Once the model of the motion control system is completed, it could be iteratively analyzed, simulated and optimized for specific scenarios, and finally, the source code or the hardware code could be automatically generated from the optimized model.

# Hardware/software co-design

There are some issues for the traditional software and hardware design methodology. Firstly, the partition between software and hardware is manually defined. Initial hardware/software partition during architectural design defines required software functions, which are possibly not optimal. Repartition means redesign, resulting in a prolonged development cycle. Secondly, the workflow is usually serial, and the software design is often after the hardware prototype design. It is not uncommon for iterating several times to finish a design. Last but not least, it lacks a unified hardware-software representation, which leads to difficulties in verifying and validating the entire system. As the increasing complexity of the system, both hardware and software are becoming more complex and less deterministic, making it much harder to accomplish design assurance. Software is not a standalone solution to hardware issues, and vice versa. A new concept is needed on the design of the hardware and software.

Hardware/software co-design is the solution to the above-mentioned issues. In order to match the diverse requirements of a product, hardware/software co-design acts as a powerful design methodology. Hardware design can inherently bring in reliable real-time performance, while software design can bring in flexibility and configurability, and hardware/software co-design is complementary in nature [33, 34]. Hardware/software co-design relies on a unified model of the target system during the product development. SystemC could be the language tool, which unified the hardware and software model. Hardware/software co-design workflow is illustrated in Figure 11. Hardware means space, and software means time. Hardware/software partition is a trade-off between flexibility and performance, and between parallel and sequence. Modeling the hardware/software system during the design process is still a non-standard and nontrivial process.

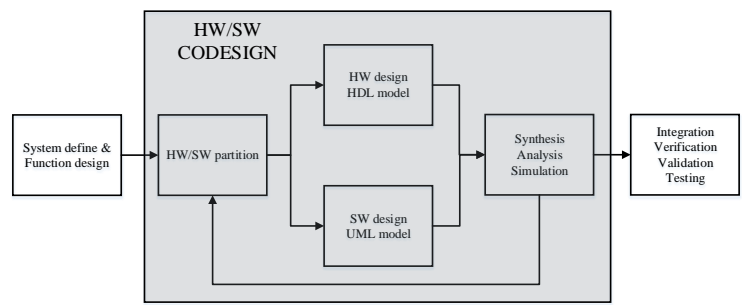

Figure 11 Hardware/software co-design workflow

Hardware/software co-design is the unique solution for extremely complex systems. For example, the Flexray unit in ASML lithography machines needs to control thousands of micro-actuators, which is a formidable task with the traditional design approach, since the scale is significant large, and the redesign and the hardware and software repartition is barely possible [35]. With hardware/software co-design, this particular controller could be implemented.

## Continuous Integration

As such complex machine like lithography tools, integration is the critical process, and there are considerable risks, such as time issues, progresses, and unexpected emerging problems. The most effective integration strategy is earlier integration, earlier exposing issues, earlier correcting, and earlier converging, which could be achieved through continuous integration.

Firstly, continuous integration could promote the integration of system model, and accelerate the top-level system design. System model integration is the first step for all integration. The model of the PCS is composed of subsystem models, which are then composed of small unite models. With continuous integration for system model, engineers could grasp the entire system from a holistic perspective. Continuous integration for integration of system models ensures the consistent of the models, guarantees the overall model optimization at the systematic level, and boosts the top-level system design and lower the risks.

Secondly, continuous integration is the best practice means for software integration and it has been successfully applied in software development [36]. Continuous integration brings multiple benefits to software integration: i) implement earlier integration, avoiding long and tense integration; ii) catch issues early and fix them early; iii) spend less time debugging and more time adding features; iv) reduce integration problems and allow engineers to deliver software more rapidly. Lithography control software, whose hierarchy is complex and the lines of code are huge, involves considerable modules. The development cycle is controllable if continuous integration is adopted.

Thirdly, continuous integration is also applicable for hardware system integration. Under the framework of hardware system, continuous integration focuses on integrating the hardware as earlier as possible and integrating all the layers from the entire system perspective, even partially for individual components, implementing steady incremental integration, which could decrease the surprises when all the hardware put together. Continuous integration for hardware system integration could make the hardware system integration more smoothly, and meanwhile it is beneficial for hardware/software co-design, co-simulation, and co-integration.

In summary, continuous integration helps system development teams be agile and respond to rapid requirement changes, ensures that the actual hardware and software development are in constant synchronization, and effectively advances the integration of the entire PCS.

# 5 Discussion

Migrating the photolithography controls system to the framework of CPS will take a long time, and revolutionary design approaches that favor of CPS requirements are not available at once. The traditional embedded system approach and the CPS approach will be coexisting for a long time, and they are mutual influence and development.

## 5.1 Learned from the PCS

The most important experience learned from the PCS is how to manipulate the extremely complicated distributed real-time control system, which could be summarized in the following.
1. There are lots of methods employed for the separation of concerns in the PCS, such as layering, partition, modularization, divide and conquer strategy.
2. There are technologies to deal with strictly hard real-time requirements. For

example, central master synchronization technology for multiple components synchronizing in an extremely tightly real-time requirement. For node to node real-time requirements, HSSL protocol, which employs the optical fiber as the physical layer, are designed. In most scenarios, CPS mostly deals with general connections, lacking of the hard real-time case. CPS could take advantage of the real-time technologies used in the PCS to extend its application areas.
3. Ad hoc approach does not fulfill the increasing demands on performance and throughput of PCS, and the model-based design, which is the fundamental part of the CPS, should be adopted in the development of the control system.

## 5.2 CPS's contribution to the PCS

CPS provides the necessary technological basis to facilitate the realization and corresponding automation of large-scale complex system. The research on CPS could benefit the PCS, and bring new thoughts and technical progress to make the control system meet the next generation demands.
1. Model-based methodology could promote the systematic development of the PCS. Although there are a few models in the current PCS, they are local, independent, and non-systematic. Under the framework of CPS, there is a formal workflow for the system development. Create system-level model first, then analyze, simulate, verify and validate based on the model, and all these are based on formal methodologies.
2. New technologies or research results developed from the CPS research, such as operating systems, programming languages and system level modeling languages, could directly apply or migrate to PCS.
3. The results in security and privacy research obtained from CPS also could applied to photolithography controls system.

## 5.3 The complexity

At present, we only have the general knowledge about the complexity of the PCS, such as numbers of actuators, sensors, hardware elements, connections and interdependencies, amounts of data stored, accessed, manipulated, and refined, and lines of code, which simply manifests the superficial complexity, lacking of the knowledge of quantity on complexity. The intricate relationships and interactions among those components constituted the complex system hide the true complexity. It is highly necessary to conduct research on the complexity of the PCS, to give a reasonable estimation for project planning, staffing, and budgets.

# 6 Conclusion

In the traditional approach, the development of PCS is following the model of embedded system development. As the complexity increasing, the traditional approach encounter insurmountable barriers, such as real-time requirements, increasing control objects, large scale, security, complexity, etc. Only under the CPS framework, and take advantage of the research results from the development of the CPS, the high demands of the PCS can be met. The advance of the CPS will definitely promote the development of the PCS. Meanwhile, PCS provides a representative case for the CPS research and development.